\documentclass[preprint2]{aastex63}

\usepackage{graphicx}
\usepackage{natbib}
\usepackage{booktabs}
\usepackage{amsmath, physics}

\received{April 7, 2021}
\revised{June 30, 2021}
\accepted{July 19, 2021}
\submitjournal{ApJ}

\shorttitle{Fast Radio Transients from M87}
\shortauthors{Suresh et al.}

\begin{document}

\title{An Arecibo Search for Fast Radio Transients from M87}

\correspondingauthor{Akshay Suresh}
\email{as3655@cornell.edu}

\author[0000-0002-5389-7806]{Akshay~Suresh}
\affiliation{Cornell Center for Astrophysics and Planetary Science, and Department of Astronomy, Cornell University, Ithaca, NY 14853, USA}

\author[0000-0002-2878-1502]{Shami~Chatterjee}
\affiliation{Cornell Center for Astrophysics and Planetary Science, and Department of Astronomy, Cornell University, Ithaca, NY 14853, USA}

\author[0000-0002-4049-1882]{James~M.~Cordes}
\affiliation{Cornell Center for Astrophysics and Planetary Science, and Department of Astronomy, Cornell University, Ithaca, NY 14853, USA}

\author[0000-0002-2578-0360]{Fronefield~Crawford}
\affiliation{Department of Physics and Astronomy, Franklin \& Marshall College, P.O. Box 3003, Lancaster, PA 17604, USA}



\begin{abstract}
The possible origin of millisecond bursts from the giant elliptical galaxy M87 has been scrutinized since the earliest searches for extragalactic fast radio transients undertaken in the late 1970s. Motivated by rapid technological advancements in recent years, we conducted $\simeq$ 10~hours of L-band (1.15--1.75~GHz) observations of the core of M87 with the Arecibo radio telescope in 2019. Adopting a matched filtering approach, we searched our data for single pulses using trial dispersion measures up to 5500~pc~cm$^{-3}$ and burst durations between 0.3--123~ms. We find no evidence of astrophysical bursts in our data above a 7$\sigma$ detection threshold. Our observations thus constrain the burst rate from M87 to $\lesssim 0.1$~bursts~hr$^{-1}$ above 1.4~Jy~ms, the most stringent upper limit obtained to date. Our non-detection of radio bursts is consistent with expectations of giant pulse emission from a Crab-like young neutron star population in M87. However, the dense, strongly magnetized interstellar medium surrounding the central $\sim 10^9 \ M_{\odot}$ supermassive black hole of M87 may potentially harbor magnetars that can emit detectable radio bursts during their flaring states.
\end{abstract}


\section{Introduction} \label{sec:intro}
The time-domain radio sky continues to reveal an abundance of astrophysical phenomena, accelerated by advances in instrumentation and computing capacity. Energetic fast radio transients (durations $\rm \lesssim 1~s$) such as fast radio bursts (FRBs: \citealt{Lorimer2007,Thornton2013,Cordes2019, Petroff2019, Chatterjee2021}), pulsar giant pulses (GPs: \citealt{Johnston2004}) and bright magnetar bursts \citep{Bochenek2020, CHIME2020a} offer the promise of discovery at extragalactic distances (Mpc--Gpc). \\

FRBs are millisecond-duration narrowband pulses of coherent radio emission originating outside our Galaxy. To date, over 600 FRB sources\footnote{FRB Newsletter Vol~2 Issue~6: \url{https://doi.org/10.7298/b0z9-fb71}} have been discovered, of which at least 24 have been seen to repeat. Precise arcsecond localization \citep{Chatterjee2017,Bannister2019,Prochaska2019,Ravi2019,Heintz2020,Law2020,Macquart2020,Marcote2020,Kirsten2021,Ravi2021,Fong2021} of 15 FRBs\footnote{\url{https://frbhosts.org}} to their respective host galaxies has revealed that FRB sources can reside in diverse host environments. Furthermore, the discovery of a luminous radio burst from the Galactic magnetar SGR~1935$+$2149 \citep{Bochenek2020, CHIME2020a} suggests a plausible magnetar engine for FRB emission. Characterized by a 1.4~GHz fluence of 1.5~MJy~ms at the 9~kpc distance \citep{Zhong2020} of SGR~1935$+$2149, such a burst would be easily detectable with $\simeq 120$~Jy~ms fluence at the $\sim$~Mpc distances to the nearest galaxies. FRB discoveries from the local Universe are hence necessary to bridge the luminosity scale between Galactic magnetars and FRBs. Detections of such bursts will further enable sensitive multi-wavelength follow-up to constrain models of FRB progenitors\footnote{\url{https://frbtheorycat.org}} \citep{Platts2019}. \\

While FRBs are of extragalactic origin, pulsar GPs constitute the most luminous Galactic radio transients at sub-millisecond timescales. First noted in the Crab pulsar PSR~J0534$+$2200 \citep{Staelin1968} and studied extensively \citep{Lundgren1995,Cordes2004,Karuppusamy2010,Karuppusamy2012,Mickaliger2012}, GPs are typically identified as short duration~($\lesssim$ ms), narrow-phase emission comprised of nanosecond-duration shot pulses \citep{Hankins2003}. GPs frequently exhibit power-law amplitude statistics \citep{Bhat2008}, unlike general pulsar single pulses \citep{Burke-Spolaor2012} that often display lognormal energy distributions. \citet{Cordes2016} evaluated the detectability of radio bursts from an extragalactic population of neutron stars that emitted nanosecond shot pulses analogous to the Crab pulsar. They demonstrate that for a fluence of $\sim~1$~Jy~ms, bursts arising from an incoherent superposition of shot pulses can be detected out to distances $\lesssim$~few$\times$~100~Mpc. The detection distance gets pushed out farther for conditions more extreme than the Crab pulsar, such as in young magnetars. Studying the GP emitter PSR~J0540$–$6919 (B0540$-$69), \citet{Geyer2021} observed band-limited flux knots analogous to that seen in FRBs. However, unlike some repeating FRBs \citep{Hessels2019,Fonseca2020}, these GPs reveal no distinct sub-pulses that drift downwards in radio frequency with increasing arrival time.\\

Hosting a $\rm M \simeq 6.5 \times 10^9 \ M_{\odot}$ supermassive black hole (SMBH;  \citealt{EHT_VI_2019}), the giant elliptical galaxy M87 within the Virgo cluster has been a popular target in past surveys for pulsed radio emission \citep{Linscott1980, Hankins1981, McCulloch1981, Taylor1981}. Akin to the Galactic Center \citep{Dexter2014}, rapid star formation near the SMBH of M87 likely yields a significant magnetar population. \citet{Michilli2018} argue that a young neutron star embedded in a strongly magnetized plasma such as that near a black hole or a supernova remnant may explain FRB~121102 (the first discovered repeating FRB: \citealt{Spitler2014, Scholz2016, Chatterjee2017}) and its large, dynamic rotation measure ($\rm |RM| \sim 10^5~rad~m^{-2}$). While the RM of FRB~121102 is unusually large among FRBs with measured RMs (typical $\rm |RM| \simeq$ 10--500~rad~m$^{-2}$, \citealt{Petroff2019}), it is comparable to that observed for the Galactic Center magnetar PSR~J1745$-$2900 ($\rm |RM| \simeq 6.6 \times 10^4~rad~m^{-2}$, \citealt{Eatough2013}). The dense, magneto-ionic interstellar medium (ISM) at the core of M87 represents a possible host for FRB~121102 and PSR~J1745$-$2900 analogs. \\

\begin{deluxetable*}{cCcccC}
\tablecaption{Log of M87 observations performed with the Arecibo radio telescope. Including overheads, each session lasted 3 hours, with varying on-source times per session. We observed a bright test pulsar for 3 minutes at the start of each session to verify proper operation of telescope electronics. \label{tab1}}
\tablewidth{0pt}
\tablehead{
\colhead{Session} & \colhead{Start MJD} & \colhead{Test pulsar} & \colhead{Net on-source time} & \colhead{Usable frequency band} & \colhead{$\mathcal{F}_{\rm min}$\tablenotemark{a}} \\
\colhead{(number)} & \colhead{(topocentric)} & \colhead{} & \colhead{(hours)} & \colhead{(GHz)} & \colhead{(Jy ms)}
}
\startdata
1 & 58505.32 & J1136$+$1551 & \phn \phn 0.60 \tablenotemark{b} & 1.15--1.50 & 1.8 \\
2 & 58506.32 & J1136$+$1551 & 2.50 & 1.15--1.75 & 1.4 \\
3 & 58507.31 & J1136$+$1551 &  \phn \phn 0.63 \tablenotemark{b} & 1.15--1.75 & 1.4 \\
4 & 58508.31 & J1239$+$2453 &  \phn  \phn 2.05  \tablenotemark{b} & 1.15--1.75 & 1.4 \\
5 & 58546.21 & J1136$+$1551 & 2.40 & 1.15--1.75 & 1.4 \\
6 & 58547.20 & J1136$+$1551 & 2.50 & 1.15--1.75 & 1.4
\enddata
\tablenotetext{a}{Minimum detectable fluence computed in accordance with Equation~\ref{eqn1} for a flat-spectrum, band-filling, boxcar-shaped pulse of width 1~ms.}
\tablenotetext{b}{Prolonged data dropouts occurred during Sessions~1, 3 and 4, restricting us from reaching our target exposure time of $\rm \simeq 2.5~hours$ per session.}
\vspace{-8mm}
\end{deluxetable*}

Intending to detect dispersed single pulses, we targeted the core of M87 with the William E. Gordon Arecibo radio telescope. Similar targeted searches for extragalactic radio bursts have previously been attempted in the direction of several galaxies \citep{McLaughlin2003, Bhat2011, Rubio-Herrera2013, vanLeeuwen2020}, including the nearby galaxies M31 and M33. \\

Section~\ref{sec:obs} describes our observing setup. We detail our data analysis methods and results in Section~\ref{sec:methods}. In Section~\ref{sec:disc}, we evaluate the significance of our results in the context of potential neutron star populations in M87. Finally, we conclude and summarize our study in Section~\ref{sec:summary}.

\section{Observations} \label{sec:obs}
Radio pulsars are steep-spectrum sources ($S_{\nu} \propto \nu^{-1.4 \pm 1.0}$, \citealt{Bates2013}), emitting greater pulse-averaged flux density ($S_{\nu}$) at lower radio frequencies ($\nu$).  As radio pulses traverse the astrophysical plasma along our lines of sight to their sources, they get dispersed (pulse arrival times $\propto \nu^{-2}$ for cold plasma dispersion) and scattered (pulse broadening time scale, $\tau_{\rm sc} \propto \nu^{-4} \rm \ or \ \nu^{-4.4}$ for Kolmogorov scattering). Optimal pulsar detection requires a suitable trade-off between the weakening pulsar emission at high radio frequencies ($\gtrsim$~10~GHz), and the growing, deleterious propagation effects at low radio frequencies ($\lesssim$~700~MHz). Large-scale pulsar surveys \citep{Manchester2001,Cordes2006,Keith2010,Barr2013,Keane2018} have hence, often been performed at 1--2~GHz, i.e., ``L--band.'' In contrast, FRB spectra are band-limited, and show no preference for a specific observing frequency. Allowing for both FRB- and pulsar-like burst spectra, L--band observations are well placed to enable extragalactic single pulse discovery from the local Universe. \\
\begin{figure}[t!]
\includegraphics[width=0.48\textwidth]{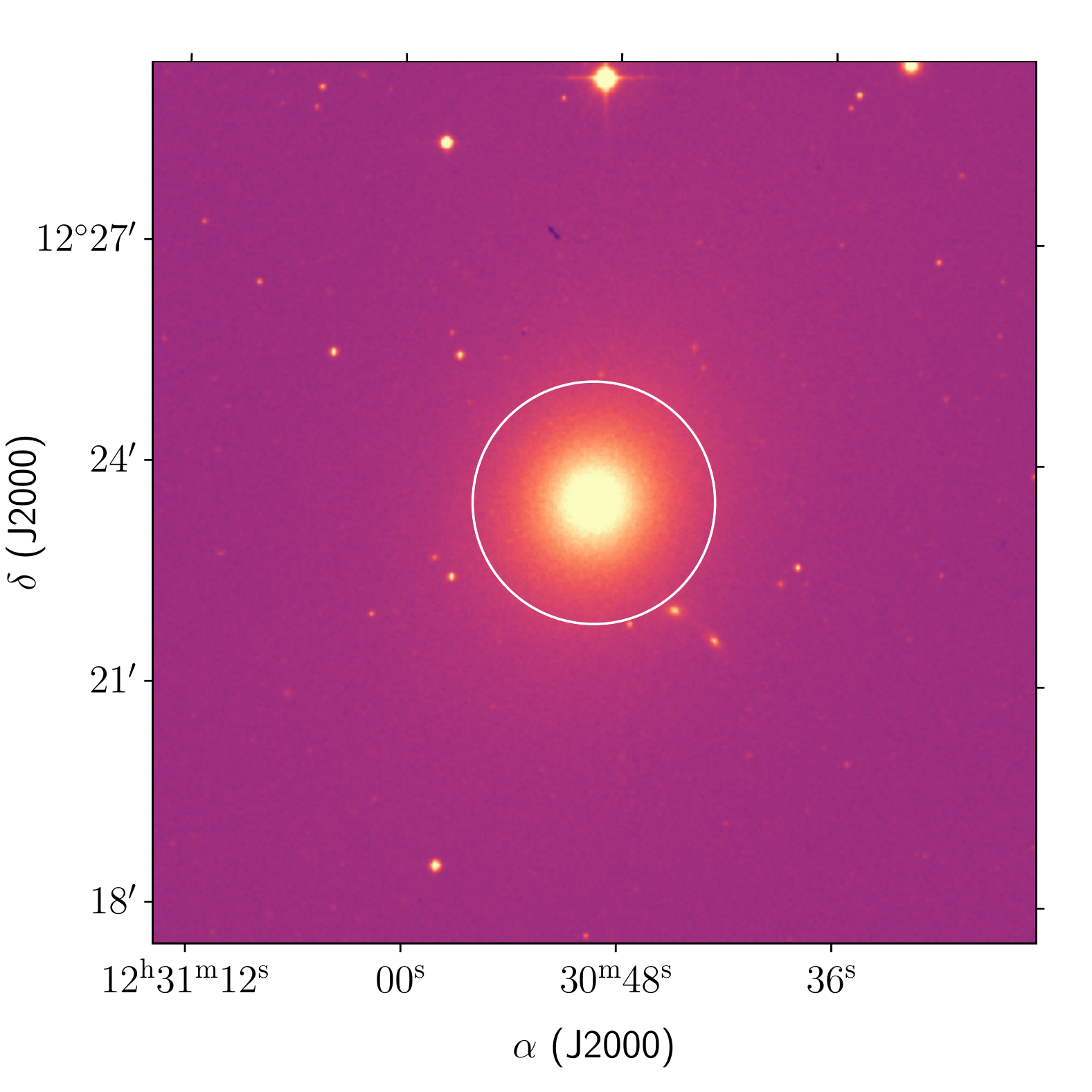}
\caption{Our L-band Arecibo beam (white circle) of HPBW $\simeq 3\farcm3$ overlaid on a Digitized Sky Survey-2-red image centered on M87. The coordinates of our pointing center are $\alpha{\rm (J2000)}= 12^{\rm h}30^{\rm m}49\fs40$ and $\delta{\rm (J2000)} = +12^{\circ} 23 \arcmin 28\farcs03$. \label{fig1}}
\end{figure}

Hunting for outbursts from the SMBH of M87, \citet{Linscott1980} detected highly dispersed (dispersion measure, $\rm DM \simeq 1000$--5500~pc~cm$^{-3}$) millisecond-duration pulses at radio frequencies of 430, 606, and 1230~MHz. However, no repeat bursts were seen in subsequent follow-up efforts \citep{Hankins1981,McCulloch1981,Taylor1981} between 400--1400~MHz. Attempting to survey the core of M87 with increased sensitivity, we executed 18 hours of L-band search-mode observations with the Arecibo radio telescope. Figure~\ref{fig1} shows our Arecibo L-band beam of HPBW $\simeq 3\farcm3$, overlaid on an optical map of M87. \\

Table~\ref{tab1} summarizes our observing program, comprised of 6 sessions lasting 3 hours (overheads included) each. We began each session with a 3-minute scan of a bright test pulsar to verify proper data acquisition system functioning. To mitigate data loss from intermittent backend malfunctions, we distributed our net on-source time per session across multiple scans of different lengths. All sessions used the single-pixel L--wide receiver with the Puerto Rico Ultimate Pulsar Processing Instrument (PUPPI) backend. The final data products generated by our observations contained 1536 usable spectral channels, each with 390.625~kHz resolution. The sampling time of our data was 64~$\mu$s. \\

As indicated in Table~\ref{tab1}, persistent data dropouts occurred during sessions 1, 3 and 4, preventing us from achieving our desired exposure time of $\simeq$ 2.5~hours per session. We discarded these dropout-affected data segments from our subsequent single pulse searches. \\

We estimate the sensitivity threshold of our observations by considering a flat-spectrum, band-filling,  boxcar-shaped pulse of width $W$. The L--band system temperature at the time of our observations was $T_{\rm sys} \simeq 27 \rm \ K$. For telescope gain, $G=10 \rm \ K~Jy^{-1}$, the corresponding system-equivalent flux density is $S_{\rm sys} = 2.7 \rm \ Jy$. The galaxy M87 contributes continuum flux density, $S_{\rm M87} \simeq \rm 212.3 \ Jy$ \citep{Perley2017} at 1.4~GHz. The radiometer equation then implies a minimum detectable fluence,
\begin{align}\label{eqn1}
\mathcal{F}_{\rm min} =& {\rm (S/N_{min})} \left( S_{\rm sys} + S_{\rm M87} \right) \left( \frac{W}{2B} \right)^{1/2}\nonumber \\
\simeq & \ 1.4 \ {\rm Jy \ ms} \left( \frac{\rm (S/N)_{min}}{7}\right) \left( \frac{S_{\rm sys} + S_{\rm M87}}{\rm 215~Jy}\right) \ldots \nonumber \\
& \left( \frac{W}{\rm 1~ms} \right)^{1/2} \left( \frac{B}{\rm 600~MHz}\right)^{-1/2}.
\end{align}
Here, $B$ and $\rm (S/N)_{min}$ denote, respectively, the observing bandwidth, and the minimum signal-to-noise ratio required to claim a detection. Table~\ref{tab1} lists $\mathcal{F}_{\rm min}$ thresholds for different observing sessions assuming a $W=1 \ \rm ms$ burst detected with $\rm (S/N)_{min} = 7$. Our observations reach down to $\mathcal{F}_{\rm min} \simeq 1.4 \ \rm Jy~ms$, about 6 times deeper than previous targeted searches \citep{Hankins1981, McCulloch1981, Taylor1981} for radio pulses from M87. For comparison, the commensal ALFABURST experiment \citep{Foster2018} at Arecibo with $B \approx 56$~MHz would have attained $\mathcal{F}_{\rm min} \simeq 4.6$~Jy~ms, i.e., a factor of $\simeq 3$ above our sensitivity limit.

\section{Methods and Results} \label{sec:methods}
\begin{figure*}[ht!]
\includegraphics[width=.48\textwidth]{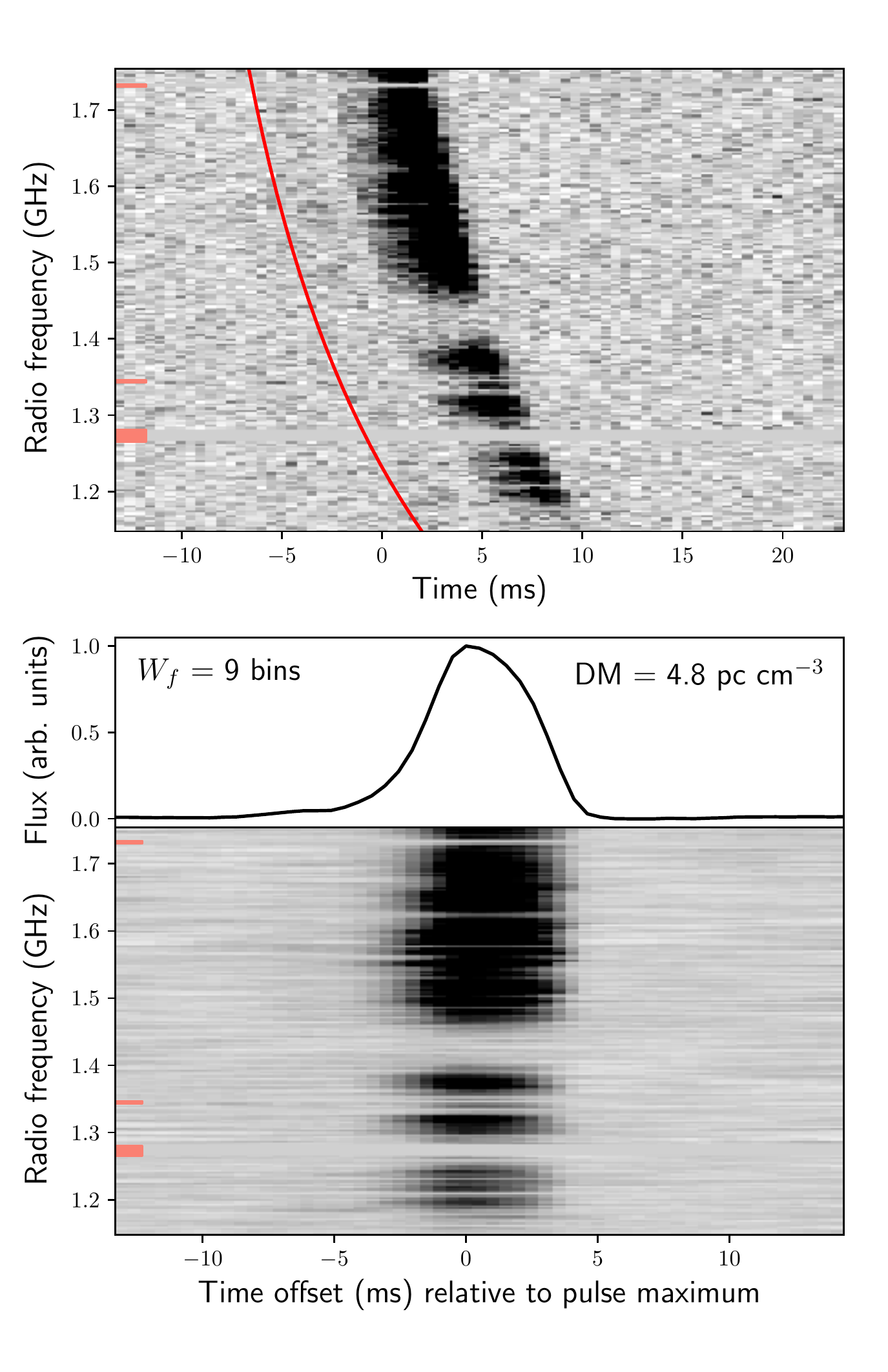}\quad
\includegraphics[width=.48\textwidth]{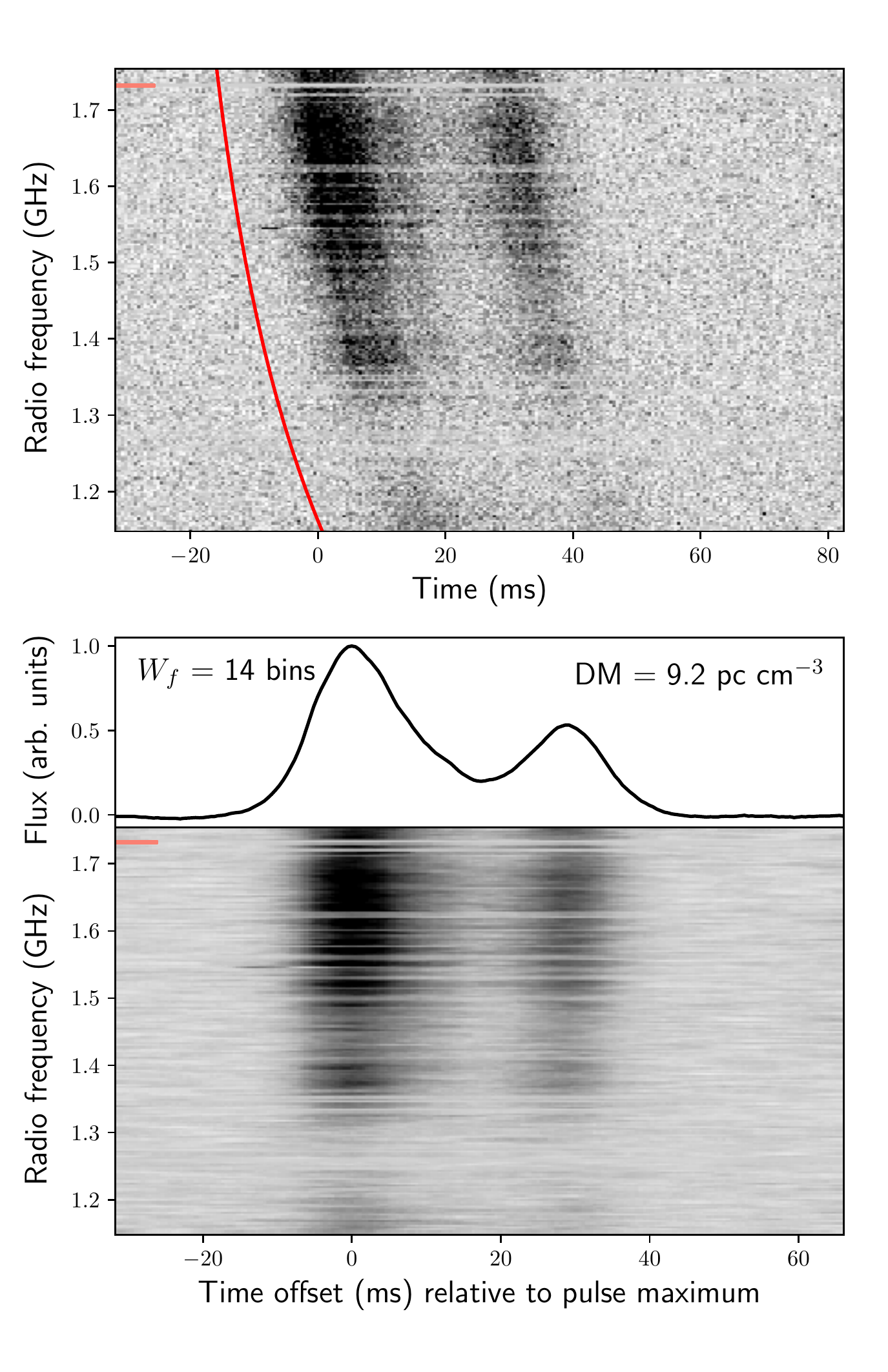}
\caption{Single pulse detections of test pulsars J1136$+$1551 (left column) and J1239$+$2453 (right column) during observing sessions 2 and 4. The top panels depict non-dedispersed dynamic spectra, block-averaged to 512~$\mu$s time resolution and 3~MHz spectral resolution. The bottom panels show dedispersed data products (dynamic spectra and time series) after convolution with their respective S/N-maximizing temporal boxcar filters. The bottom panels also quote the pulse DM and the S/N-maximizing temporal boxcar filter width ($W_f$). The short orange dashes at the left edges of all dynamic spectra represent channels flagged by our RFI excision procedure. The red curves in the top panels illustrate  $\nu^{-2}$ dispersion curves corresponding to the pulsars' DMs. \label{fig2}}
\end{figure*}
Conventional searches for dispersed pulses typically involve matched filtering of dedispersed time series with template filters of various widths. However, the ubiquitous presence of radio frequency interference (RFI) in dynamic spectra (radio frequency-time plane) often complicates such searches. We discuss our RFI excision procedure in Section~\ref{sec:RFI}. Following RFI masking, we illustrate data integrity through our test pulsar detections in Section~\ref{sec:testpulsar}. Since the true DM of a radio burst is unknown prior to discovery, dynamic spectra need to be dedispersed over a range of trial DMs. These dedispersed dynamic spectra, one per trial DM, are then summed over radio frequency to produce dedispersed time series for single pulse searching. Sections~\ref{sec:dedispersion} and \ref{sec:spsearch} describe our dedispersion plan and single pulse search methodology respectively.

\subsection{RFI Excision}\label{sec:RFI}
Informed by Arecibo-specific RFI mitigation performed by \citet{Lazarus2015}, we used the {\tt rfifind} module of the pulsar search software {\tt PRESTO} \citep{PRESTO} to operate on 1-second sub-integrations of data. For each 1-second block in every frequency channel, {\tt rfifind} computes two time-domain statistics, namely the block mean and the block standard deviation. A Fourier-domain statistic, i.e., the maximum of the block power spectrum, is also calculated. Blocks with one or more statistics that deviate significantly from the means of their respective distributions are labeled as RFI. For the time-domain statistics, we adopted a flagging threshold of 5 standard deviations from the distribution mean. The corresponding threshold for the Fourier-domain statistic was 4 standard deviations from the mean. \\

To mask RFI, the ensuing set of flagged blocks were replaced by median bandpass values of that time range. Time integrations containing over $50\%$ flagged channels were masked completely. Likewise, channels with at least $20\%$ flagged blocks were entirely replaced by zeros. All flagging thresholds chosen in our study were conservative choices based on visual inspection of short data segments and parameter estimates from \citet{Lazarus2015}. \\

To remove broadband baseline fluctuations, we applied a zero-DM filter to subtract the mean over channels from each time slice in the masked, non-dedispersed dynamic spectrum. \citet{Eatough2009} investigated the sensitivity loss from zero-DM filtering for boxcar single pulse detection in the Parkes Multi-beam Pulsar Survey ($\nu = 1.4$~GHz, $B \approx 288$~MHz, \citealt{Manchester2001}). While ${\rm DM} = 0$~pc~cm$^{-3}$ signals get completely eliminated, boxcar pulses with widths, $W \lesssim 9$~ms, can be detected with $\gtrsim 90\%$ sensitivity at ${\rm DM} \simeq 100$~pc~cm$^{-3}$. The detection sensitivity to broader pulses increases further at higher DMs. \\

Implementing the above RFI excision process, the prominent signals masked out in our data include  intermittent, narrow-band RFI between 1.26--1.28 and 1.72--1.73~GHz. In summary, up to 95--100$\%$ of our observing bandwidth was usable every session. 

\subsection{Test Pulsar Verification}\label{sec:testpulsar}
As listed in Table~\ref{tab1}, our observing program included 3-minute scans of the bright test pulsars J1136$+$1551 (B1133$+$16) and J1239$+$2453 (B1237$+$25). To detect the periodicity of these pulsars, we first dedispersed our pulsar dynamic spectra to their respective known pulsar DMs. Using the {\tt prepfold} routine of {\tt PRESTO}, we then ran a blind folding search for periodic pulsations in these dedispersed data. In doing so, we recovered pulsar rotational periods and average pulse profiles that were consistent with previously published results \footnote{\url{http://www.atnf.csiro.au/research/pulsar/psrcat}} \citep{Manchester2005}. \\

\begin{deluxetable*}{CCCCC}
\tablecaption{Dedispersion and single pulse search plan. We used boxcar filters of widths 1, 2, 3, 4, 6, 9, 14, 20, and 30 bins to perform matched filtering for single pulses in dedispersed time series. \label{tab2}}
\tablewidth{0pt}
\tablehead{
\colhead{DM range} & \colhead{DM step size} & \colhead{No. of trial DMs} & \colhead{Downsampling factor} & \colhead{Range of boxcar widths}  \\
\colhead{(pc~cm$^{-3}$)} & \colhead{(pc~cm$^{-3}$)} & \colhead{} & \colhead{} & \colhead{(ms)}
}
\startdata
\phn \phn 0.0-693.6 & 0.2 & 3468 & \phn 4 & 0.256-7.68 \phn \\
\phn 693.6-1183.2 & 0.3 & 1632 & \phn 8 & 0.512-15.36 \\
1183.2-2203.2 & 0.5 & 2040 & 16 & 1.024-30.72 \\
2203.2-4243.2 & 1.0 & 2040 & 32 & 2.048-61.44 \\
4243.2-5500.2 & 3.0 & \phn 419 & 64 & \phn 4.076- 122.88
\enddata
\vspace{-8mm}
\end{deluxetable*}
In addition to periodicity confirmation, we searched our test pulsar data for single pulses. To do so, we dedispersed our pulsar data over trial DMs ranging from 0~pc~cm$^{-3}$ to 100~pc~cm$^{-3}$, with a DM grid spacing of 0.4~pc~cm$^{-3}$. We then block-averaged our dedispersed time series to 512~$\mu$s resolution, and searched these time series for single pulses using a matched filtering approach. We accomplished our single pulse searches using the {\tt single$\_$pulse$\_$search.py} module of {\tt PRESTO}, which convolves an input time series with boxcar filters of various widths. We considered boxcar filter widths of 1, 2, 3, 4, 6, 9, 14, 20, and 30 bins in our burst search analysis.  \\ 

Let $\rm (S/N)_{mf}$ denote the S/N of a single pulse candidate in the convolution of its dedispersed time series with a boxcar matched filter. Setting $\rm (S/N)_{mf} \geq 10$ as the detection criterion, we successfully detected dispersed pulses in all test pulsar scans.  Figure~\ref{fig2} shows single pulse detections of pulsars J1136$+$1551 and J1239$+$2453 during observing sessions~2~and~4 respectively.  The pulse from J1136$+$1551 is from only one of the two primary components seen in the average profile, while for J1239$+$2453, the pulse in the top panel shows emission in several of the five profile components. Matched filtering smears some of this structure in the bottom panel. 

\subsection{Dedispersion Plan}\label{sec:dedispersion}
\begin{figure*}[ht!]
\includegraphics[width=\textwidth, trim={0.25cm 0.5cm 1.4cm 1cm}, clip]{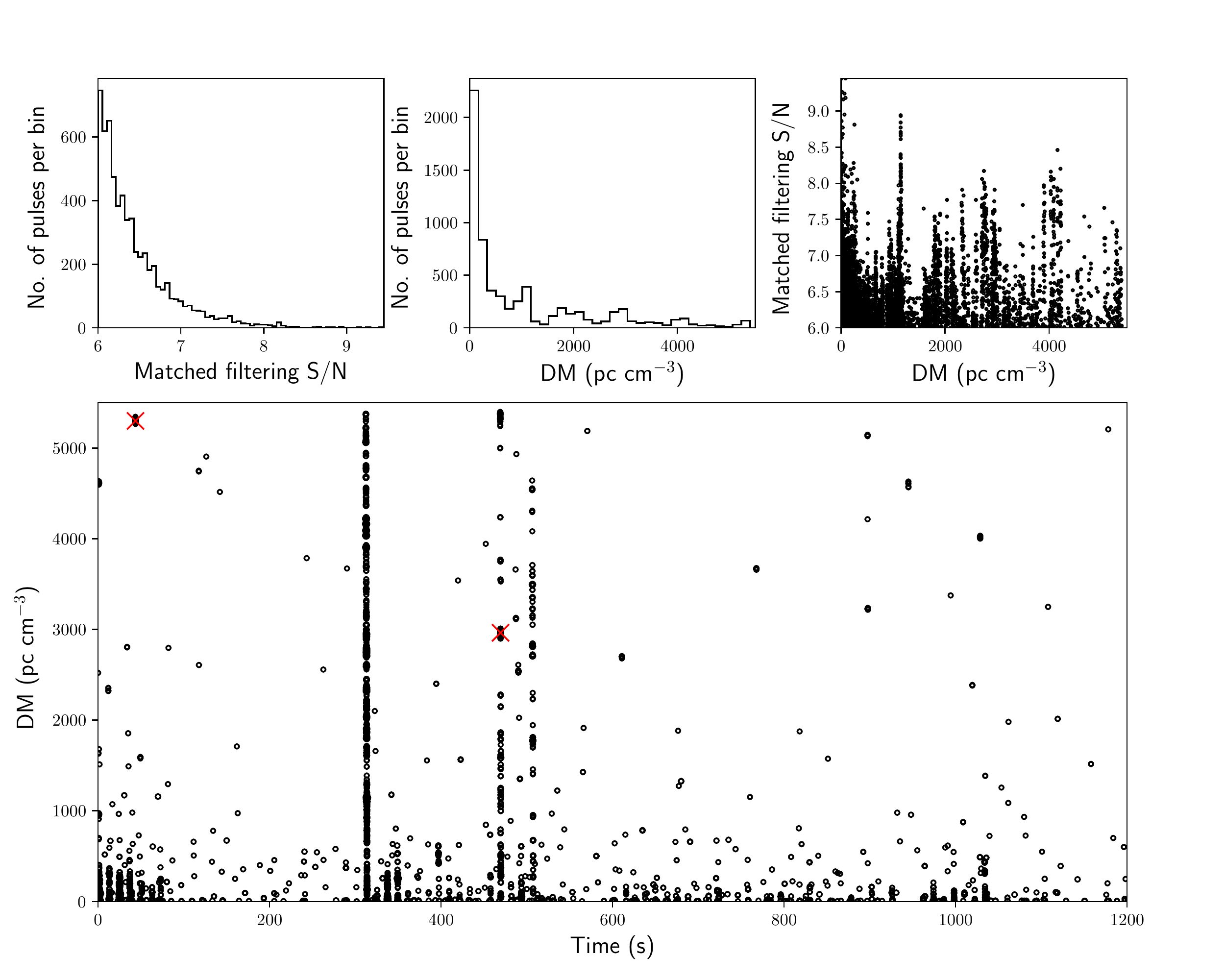}
\caption{A sample single pulse search output from a 20-minute scan of M87 during session~6. Single pulse candidates with matched filtering S/N, $\rm (S/N)_{mf} \geq 6$, are plotted in this figure. The top left and top middle panels illustrate histograms of $\rm (S/N)_{\rm mf}$ values and candidate DMs respectively. The top right panel plots $\rm (S/N)_{\rm mf}$ against candidate DMs. The bottom panel shows a scatter plot of single pulse candidates in the DM-time plane. Each circular marker represents a single pulse candidate, with the radius of the circle proportional to $\rm (S/N)_{\rm mf}$. The two red crosses mark candidates whose smoothed, dedispersed dynamic spectra are shown in Figure~\ref{fig4}. \label{fig3}}
\end{figure*}
Looking to find possible repeats of the \citet{Linscott1980} bursts, we dedispersed our M87 data out to $\simeq$ 5500~pc~cm$^{-3}$. Table~\ref{tab2} summarizes our dedispersion plan, which attempts to optimize various contributions to pulse broadening. For a pulse of intrinsic width~$W$, its effective width \citep{Cordes2003} in a dedispersed time series is 
\begin{align}\label{eqn2}
W_{\rm eff} = \left( W^2 + t_{\rm samp}^2 + \tau_{\rm sc}^2  +  t_{\Delta \nu}^2 + t_{\rm chan}^2 + t_{\rm BW}^2  \right)^{1/2}.
\end{align}
Here, $t_{\rm samp}$ is the sample interval, and $\tau_{\rm sc}$ is the scatter-broadening time scale. For channel bandwidth $\Delta \nu$,  $t_{\Delta \nu} \sim (\Delta \nu)^{-1}$ is the receiver filter response time. At radio frequency $\nu$, the intrachannel dispersive smearing is
\begin{align}\label{eqn3}
t_{\rm chan} \simeq 8.3~\mu{\rm s} \left(\frac{{\rm DM}_{\rm pc~cm^{-3}} \ \Delta \nu_{\rm MHz}}{\nu_{\rm GHz}^3} \right).
\end{align}
The use of a finite DM step size ($\delta \rm DM$) for dedispersion introduces a residual broadband dispersive delay given by
\begin{align}\label{eqn4}
t_{\rm BW} \simeq 8.3~\mu{\rm s} \left(\frac{{\rm \delta DM}_{\rm pc~cm^{-3}} \ B_{\rm MHz}}{\nu_{\rm GHz}^3} \right),
\end{align}
where $B$ is the observing bandwidth. Since $\tau_{\rm sc}$ cannot be corrected in practice, we neglect it when devising our dedispersion plan. Therefore, the net optimizable contribution to the effective pulse width is
\begin{align}\label{eqn5}
t_{\rm tot} = \left(  t_{\rm samp}^2 + t_{\rm chan}^2 + t_{\rm BW}^2 \right)^{1/2}.
\end{align}
\begin{figure*}[th!]
\includegraphics[width=.48\textwidth]{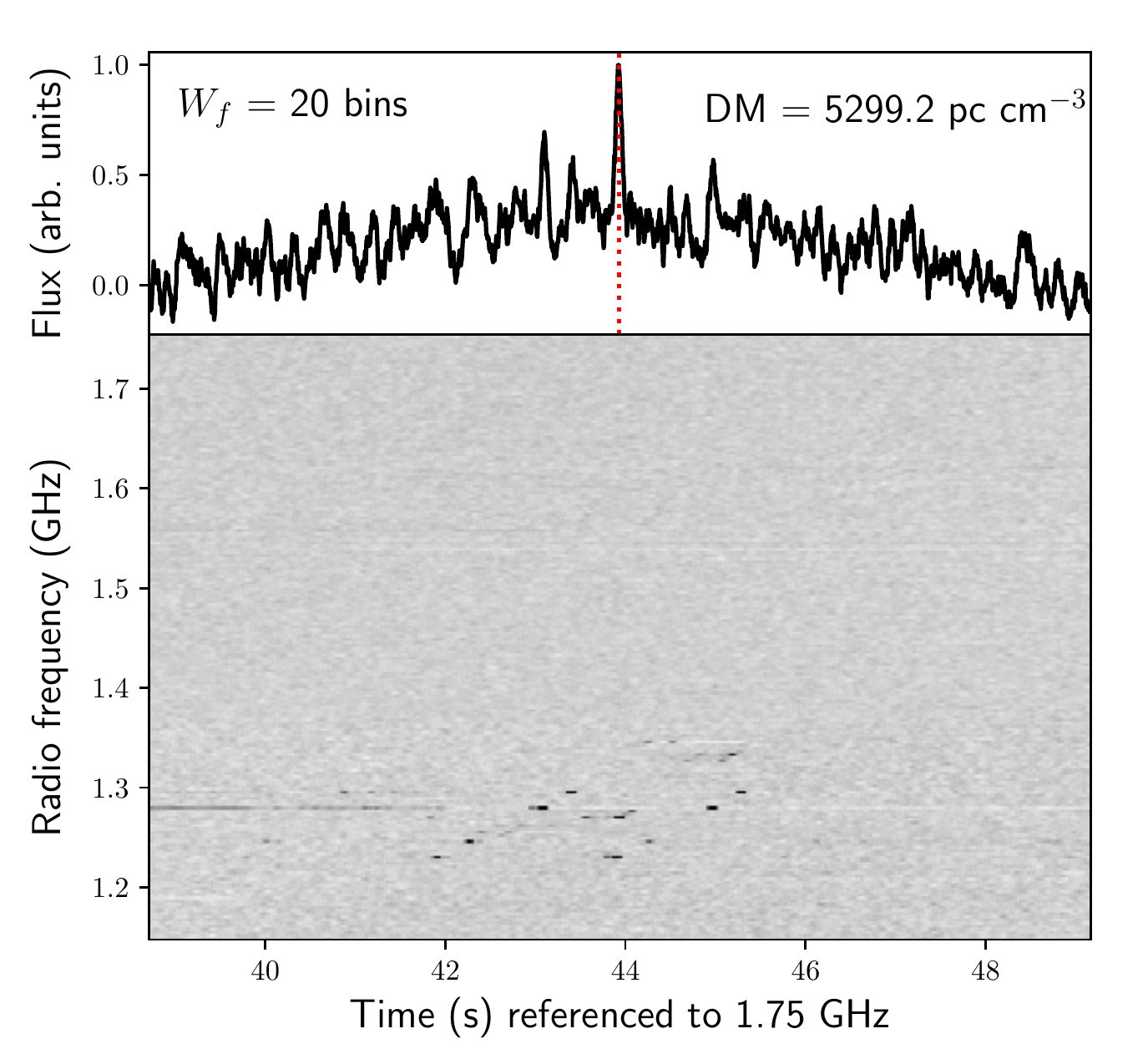}\quad
\includegraphics[width=.48\textwidth]{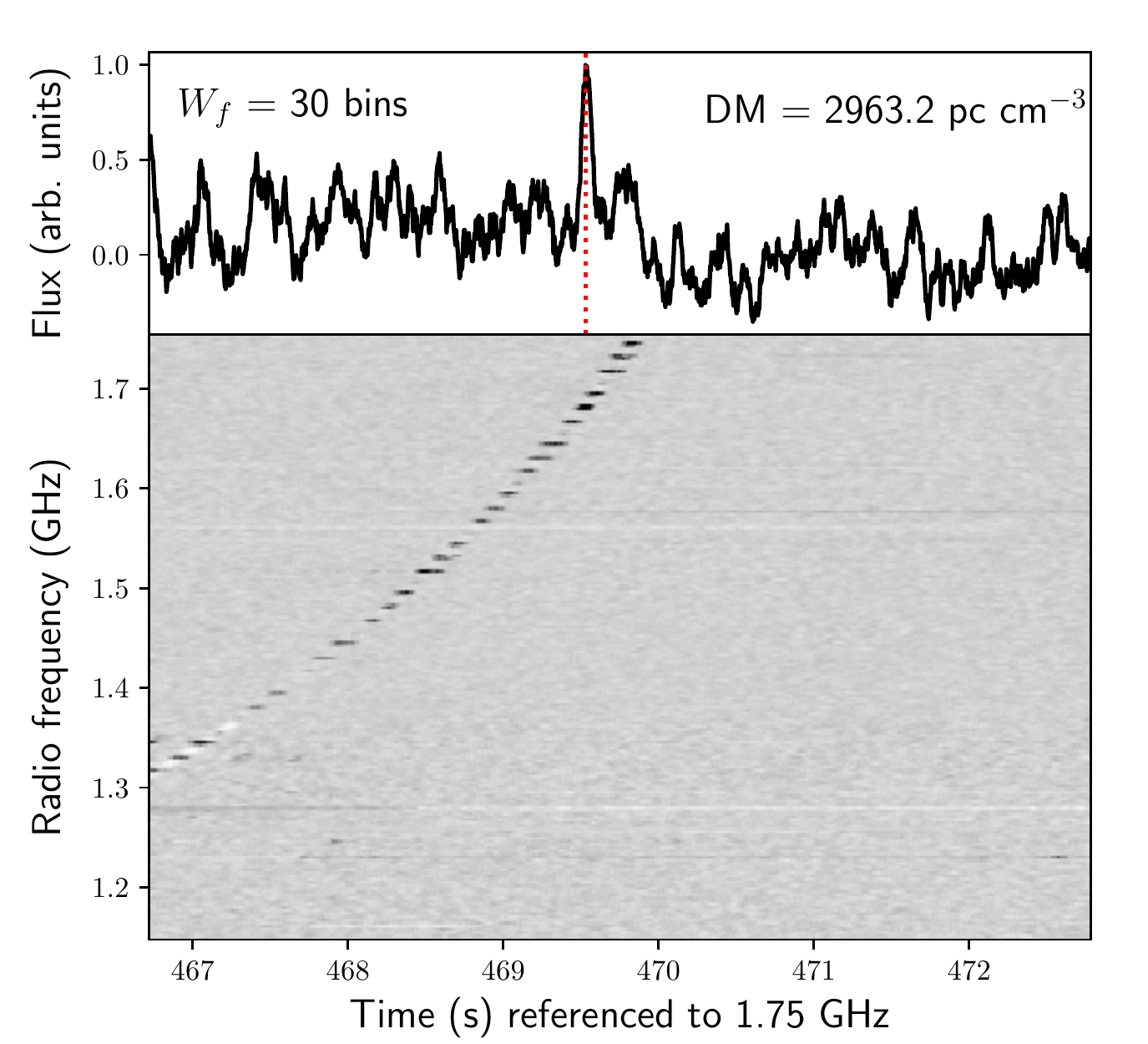}
\caption{Smoothed, dedispersed dynamic spectra (bottom subplot of each panel) and time series (top subplot of each panel) of RFI signals missed by our RFI excision method (Section~\ref{sec:RFI}) and marked with red crosses in Figure~\ref{fig2}. The red vertical dotted lines in the top subplots represent candidate peak times in their respective dedispersed time series. The top subplots also quote the candidate DMs and the S/N-maximizing temporal boxcar filter widths ($W_f$) used for smoothing. The reference frequency for dedispersion is the top of the observing band at 1.75~GHz.\label{fig4}}
\end{figure*}
With increasing DM, $t_{\rm chan}$ grows and dominates $t_{\rm tot}$. To minimize computational cost, we downsampled our dedispersed time series via block-averaging, and increased $\delta \rm DM$ at higher DMs. Table~\ref{tab2} lists temporal downsampling factors and $\delta \rm DM$ values for various trial DM ranges in our study. \\

Through matched filtering, we incur negligible loss of sensitivity in our burst searches \citep{Keane2015}. For RFI-cleaned data, the finite DM grid explored in our study then determines our survey completeness. Specifically, for $\delta {\rm DM} \simeq 1$~pc~cm$^{-3}$, $t_{\rm BW}$ limits burst detection for $W \lesssim 2$~ms. Hence, we chose downsampling factors in Table~\ref{tab2} that provide optimal sensitivity to burst durations in different DM ranges.

\subsection{Single Pulse Searching}\label{sec:spsearch}
Following the single pulse search methodology described in Section~\ref{sec:testpulsar}, we ran burst search analyses on our M87 data. Again, we operated with boxcar filters of widths 1, 2, 3, 4, 6, 9, 14, 20, and 30 bins for matched filtering. Table~\ref{tab2} lists the boxcar filter durations used for DM ranges with distinct downsampling factors. Our boxcar filters span widths, $W_f \simeq$ 0.3--8~ms at the lowest DMs to $W_f \simeq$ 4--123~ms at the highest trial DMs covered in our study. \\

Figure~\ref{fig3} shows a sample single pulse search output from a 20-minute scan of M87 during session~6. Real astrophysical bursts are expected to manifest as localized spindles with non-zero central DMs in the DM-time plane. To verify the presence of such signals in our data, we visually inspected dynamic spectra of all promising candidates with matched filtering S/N, $\rm (S/N)_{\rm mf} \geq 7$. Figure~\ref{fig4} illustrates dedispersed dynamic spectra of two such candidates that were examined. \\  

\begin{deluxetable*}{lCCCCC}
\tablecaption{Comparison of fluence thresholds ($\mathcal{F}_{\rm min}$) and burst rates/limits ($\mathcal{R}$) across surveys for pulsed radio emission from M87. \label{tab3}}
\tablewidth{0pt}
\tablehead{
\colhead{Study} & \colhead{Radio frequency} & \colhead{$\mathcal{F}_{\rm min}$\tablenotemark{a}} & \colhead{$N_{\rm pulses}$} & \colhead{On-source time} & \colhead{$\mathcal{R} (\mathcal{F} \geq \mathcal{F}_{\rm min})$} \\
\colhead{} & \colhead{(MHz)} & \colhead{(Jy~ms)} & \colhead{} & \colhead{(min.)} & \colhead{(bursts hr$^{-1}$)}
}
\startdata
This work\tablenotemark{b} & 1400 \phn & 1.4 & 0 & 605 \phn \phn & \leq \phn~0.1 \phn \phn \\
This work + $\mathcal{R} (> \mathcal{F}) \propto \mathcal{F}^{-1}$~\tablenotemark{c} & 1400 \phn & \phn 9 & 0 & 605 \phn \phn & \leq~0.02 \phn \phn \\
\citet{Hankins1981}\tablenotemark{d} & 430 & 105 \phn & 0 & 2 & \leq \phn \phn 30 \phn \phn \\
& 606 & 138 \phn & 0 & 1.1~\phn & \leq \phn \phn 55 \phn \phn \\
& 1020 \phn& 120 \phn & 0 & 0.4~\phn & \leq \phn 150 \phn \phn \\
& 1400 \phn & \phn 9 & 0 & 120 \phn \phn & \leq \phn~0.5 \phn \phn \\
\citet{McCulloch1981}\tablenotemark{d} & 1400 \phn & 15 & 0 & 60 \phn & \leq \phn \phn \phn 1 \phn \phn \\
\citet{Taylor1981}\tablenotemark{d} & 606 & 77 & 0 & 6.7~\phn & \leq \phn \phn \phn 9 \phn \phn \\
\citet{Linscott1980} & 430 & 109 \phn & 21 \phn & 1 & 1260 \\
& 606 & 88 & 23 \phn & 1 & 1380 \\
& 1400 \phn & 42 & 22 \phn & 1 & 1320
\enddata
\tablenotetext{a}{Minimum detectable fluence corresponding to 7$\sigma$ detection of a 1~ms Boxcar-shaped pulse.}
\tablenotetext{b}{We ignore session~1 due to its marginally higher $\mathcal{F}_{\rm min}$ compared to other sessions.}
\tablenotetext{c}{$\mathcal{R} (> \mathcal{F}) \propto \mathcal{F}^{-1}$ scaling applied to facilitate comparison of $\mathcal{R}$ with that obtained by \citet{Hankins1981} in their 1400~MHz observations.}
\tablenotetext{d}{\citet{Hankins1981}, \citet{McCulloch1981}, and \citet{Taylor1981} experimented with multiple instrumental setups at each observing frequency. For a given radio frequency, we quote here $\mathcal{F}_{\rm min}$ from their most sensitive observation.}
\vspace{-8mm}
\end{deluxetable*}
To discern dispersed bursts from RFI in dynamic spectra, \citet{Foster2018} devised a set of metrics based on a prototypical pulse model. However, RFI can manifest with diverse spectro-temporal morphologies and variable signal strengths, thereby rendering the burst S/N, bandwidth, and duration as unreliable classification criteria. We therefore demanded the presence of a continuous $\nu^{-2}$ dispersive sweep and natural burst sub-structure (analogous to known FRB and GP discoveries) as litmus tests for astrophysical pulses. We also entertained the notion of DM consistency across possible repeat events with the caveat that DMs may significantly vary between burst sources in different regions of M87. Adopting the above selection criteria, our manual inspection process reveals that all candidates with $\rm (S/N)_{\rm mf} \geq 7$ can be attributed to short duration ($\simeq$ 100~ms) RFI patches that were missed by our RFI excision procedure.\\

We set $\rm (S/N)_{\rm min} = 7$ as the detection threshold for our M87 burst searches. Our non-detection of dispersed pulses in 10 hours of integration time then imposes the upper limit $\mathcal{R} \lesssim 0.1$~bursts~hr$^{-1}$ on the burst rate ($\mathcal{R}$) from M87 above $\mathcal{F}_{\rm min} \simeq 1.4$~Jy~ms, assuming a fiducial burst width of 1~ms.

\section{Discussion} \label{sec:disc}
Table~\ref{tab3} summarizes burst rates/limits derived from all known searches for radio pulses from M87. Evidently, our Arecibo observations constitute the deepest single pulse searches of M87 conducted to date. Assuming a cumulative burst fluence distribution, $\mathcal{R} (> \mathcal{F}) \propto \mathcal{F}^{-1}$, similar to that seen for FRB~121102 \citep{Law2017,Gourdji2019,Oostrum2020,Cruces2021}, our observations constrain $\mathcal{R}$ to at least a factor of 25 better than previous surveys of M87. We postulate a likely non-astrophysical origin for the \citet{Linscott1980} pulses given their inconsistency with the burst non-detection reported in more sensitive surveys of M87. In the following paragraphs, we explore the significance of our radio burst non-detection in the context of likely neutron star populations in M87. \\

The Crab pulsar, with a characteristic age of $\tau_c \simeq 1300$~years, is among the best studied GP emitters in our Galaxy. Based on a sample of $\simeq$ 13,000 Crab GPs at 1.4~GHz, \citet{Karuppusamy2010} inferred a cumulative burst rate distribution,
\begin{align}\label{eqn6}
\mathcal{R}_{\rm Crab} (>\mathcal{F}) \simeq 7~{\rm bursts~min}^{-1} \left( \frac{\mathcal{F}}{\rm 2~Jy~ms} \right)^{\alpha},
\end{align}
above $\mathcal{F} \simeq 2$~Jy~ms. Empirical values of the power-law index, $\alpha$, range from $\approx -2.5$ to $-1.3$ depending on the observation epoch and the observing frequency \citep{Mickaliger2012}. Here, we nominally adopt $\alpha=-2$ for illustration. \\

Following \citet{Cordes2016}, we extend $\mathcal{R}_{\rm Crab} (>\mathcal{F})$ to extragalactic radio pulsars and assess the detectability of Crab-like GPs from M87. Considering millisecond bursts from M87 (distance $\simeq$ 16.4~Mpc, \citealt{EHT_VI_2019}), our detection threshold of 1.4~Jy~ms corresponds to a limiting fluence of 94~MJy~ms at the 2~kpc distance \citep{Trimble1973} of the Crab pulsar. 
Equation~\ref{eqn6} then implies a negligible rate of $\simeq 2$~bursts~Gyr$^{-1}$ for Crab-like GPs from M87. GP detection from M87 therefore, entails young neutron stars capable of emitting more frequent supergiant pulses than the Crab pulsar. \\

We estimate the probable number of young pulsars in M87, starting from the Galactic canonical pulsar birth rate, $\beta_{\rm PSR,~MW} = 1.4~{\rm century}^{-1}$ \citep{Lorimer2006}. The star formation rate (SFR) in M87 is $\simeq 0.05 M_{\odot}~{\rm yr^{-1}}$ \citep{Terrazas2017}, about 38 times smaller than that of the Milky Way \citep{Chomiuk2011}. Scaling $\beta$ linearly with SFR, we expect $\leq 1$ canonical pulsar in M87 younger than the Crab pulsar. The low SFR of M87 thus renders unlikely the prospect of detecting GPs from canonical pulsars in M87. \\

Aside from canonical pulsars, alternate potential burst sources include millisecond pulsars (MSPs) prevalent in globular clusters \citep{Ransom2008}, magnetars theorized to power FRBs \citep{Lyubarsky2014,Beloborodov2017,Margalit2019,Metzger2019}, and binary neutron star mergers emitting radio precursors \citep{Sridhar2021}. M87 hosts a rich globular cluster system \citep{Strader2011}, with $\simeq 650$ globular clusters contained inside our Arecibo HPBW $\simeq 3\farcm3$ ($\approx 15.7$~kpc). Galactic pulsar surveys have thus far uncovered $\simeq 120$ millisecond pulsars in 36 globular clusters\footnote{\url{http://www.naic.edu/~pfreire/GCpsr.html}}, equating to a mean discovery rate of $\simeq 3$~MSPs per globular cluster. Extending this rate to M87 using a linear scaling with SFR, we predict at least $\simeq 50$ MSPs to be contained inside our Arecibo beam. However, single pulse detections from such objects are extraordinarily unlikely, requiring exotic systems emitting bursts $\gtrsim 10^{8}$ times more energetic than GPs from Galactic MSPs. For example, the brightest GP detected from the Galactic MSP B1937$+$21 \citep{Backer1982} has fluence, $\mathcal{F} \simeq 200$~Jy~$\mu$s \citep{McKee2019} at 1.4~GHz. Placing this burst source at the distance to M87, we observe a practically undetectable burst fluence, $\mathcal{F} \simeq 10$~nJy~ms $\sim 10^{-8} \mathcal{F}_{\rm min}$. Moreover, our survey parameters together with the lack of baseband data (raw complex voltages) render potential GP detection from MSPs unlikely.\\

Magnetar births and neutron star mergers \citep{Artale2020} are generally associated with gas-rich, star-forming regions in the Universe. But, such locations are scarce in a red elliptical galaxy like M87. Motivated by the hitherto non-detection of Galactic Center pulsars and the discovery of a single magnetar \citep{Eatough2013} at the Galactic Center, \citet{Dexter2014} suggest that strong ISM magnetic fields in the vicinity of a SMBH could boost magnetar production. However, a robust evaluation of burst detectability is difficult due to large uncertainties in intrinsic magnetar energy budgets, lengths of flaring and quiescent periods, and beaming geometries relative to our lines of sight.

\section{Summary and Conclusions} \label{sec:summary}
We executed a set of 1.15--1.75~GHz observations of the core of M87 with the Arecibo radio telescope in order to search for millisecond bursts. Our observations lasted a total of 18 hours, of which 10 hours were spent on-source. Using a matched filtering approach, we searched our data for single pulses, at trial DMs up to 5500~pc~cm$^{-3}$ with boxcar filter widths between 0.3--123~ms. Adopting a $7\sigma$ detection criterion, we report the non-detection of astrophysical bursts in our data, implying a burst rate limit $\mathcal{R} \lesssim 0.1$~bursts~hr$^{-1}$ above $\mathcal{F}_{\rm min} \simeq 1.4$~Jy~ms. Invoking $\mathcal{R}(>\mathcal{F}) \propto \mathcal{F}^{-1}$, our observations constrain $\mathcal{R}$ to at least a factor of 25 better than previous single pulse searches of M87. We suggest a non-astrophysical origin for the \citet{Linscott1980} burst discoveries based on their non-confirmation in more sensitive subsequent surveys of M87. \\

We evaluated the significance of our radio burst non-detection in the context of different neutron star populations in M87. Millisecond pulsars are too weak to yield detectable emission at extragalactic distances, and the low star formation rate of M87 renders unlikely the existence of a significant Crab-like, GP-emitting pulsar population. Magnetars may however reside in the dense magneto-ionic medium near the SMBH of M87. Such magnetars may emit sufficiently energetic radio pulses for detection during their active phases. We encourage high sensitivity, multi-epoch observations of M87 to detect possible magnetar radio bursts, if they are favorably beamed towards our line of sight.

\acknowledgments
AS thanks Scott~M.~Ransom for helpful software-related discussions. AS, SC and JMC acknowledge support from the National Science Foundation (NSF AAG$-$1815242). SC, JMC and FC are members of the NANOGrav Physics Frontiers Center, which is supported by the NSF award PHY$-$1430284. \\

The Arecibo Observatory was a facility of the National Science Foundation operated under cooperative agreement by the University of Central Florida and in alliance with Universidad Ana G. Mendez, and Yang Enterprises, Inc. The Arecibo observations presented here were gathered as part of program P3315, PI: A.~Suresh. \\

This work used the Extreme Science and Engineering Discovery Environment (XSEDE) through allocation PHY200054, which is supported by National Science Foundation grant number ACI$-$1548562. Specifically, it used the Bridges system, which is supported by NSF award number ACI$-$1445606, at the Pittsburgh Supercomputing Center (PSC). AS acknowledges the XSEDE and PSC staff for their timely and helpful responses to queries. 

%

\vspace{5mm}
\facility{Arecibo, XSEDE \citep{XSEDE, Bridges}.}


\software{Astropy \citep{Astropy},
                 NumPy \citep{NumPy},
                 Matplotlib \citep{Matplotlib},
                 PRESTO \citep{PRESTO},
                 Python~3 (\url{https://www.python.org}), 
                 SciPy \citep{SciPy}.
                }

\bibliography{references}
\bibliographystyle{aasjournal}

\end{document}